\documentclass{appolb}
\usepackage{epsfig}

\begin{document}

\def\tj{trajectory }
\def\tjn{trajectory}
\def\n{$n$ }
\def\nn{$n$}
\def\cn{curve}
\def\c{curve }
\title{ Curves and The  Photon 
\thanks{Presented at PHOTON 2005, Warsaw University, Sept. 2005}
}
\author{L. Stodolsky
\address{Max-Planck-Institut f\"ur Physik 
(Werner-Heisenberg-Institut),
F\"ohringer Ring 6, 80805 M\"unchen, Germany}
}
\maketitle
\begin{abstract}
The study of the number of photons leads to a new way of
characterizing curves and to a
novel integral invariant over curves.
\end{abstract}
  
\section{Introduction}
It's always surprising how far one good idea can go. And while
we're here
celebrating the 100th birthday of the photon, I would like to add
one more
fillip on that good idea. This time verging into
almost pure mathematics; the photon leads us to a new way of
characterizing curves and to a novel integral invariant over
curves.

 In classical radiation theory practically all there is to
calculate is the
energy radiated; or perhaps occasionally a frequency spectrum or a
polarization.
 But with the advent of the photon there is a new quantity in the
 radiation field: just the plain  number of photons. This is
 an ordinary real number, evidently dimensionsless.  And
independent of
the Lorentz frame; while changing your velocity can change the
energy of a photon, it can't change one photon into two photons.
 If it is well-defined there ought to be some nice simple formula
for it.
``But aha'', you will say, ``that's just the trouble, it's not well
defined.
There's the notorious infra-red catastrophe, where as we all know,
the number
of photons radiated is  infinite--even while the radiated energy
remains finite.''

Correct. But nevertheless, as  I was surprised to realize, there is
a general
class of situations where the infra-red catastrophe is averted.

\section{Finite Number of Photons}
Let $n$ be the number of photons radiated by a charge following
some given trajectory.
Let  the initial
and final velocities be  exactly the same---let us  ``identify'' 
them. In this case \n can in fact be finite. This is because
in a scattering process the infra-red catastrophe  arises from the
infinite flight paths for the
incoming and outgoing particle, as the infinite
range field is shaken off or re-constituted. Or in terms of Fourier
transforms, zero frequency results from infinite time. When the 
re-constituting is not necessary
because the infinite paths are identical, or when the infinite
times don't occur, the number of photons can be
finite. Naturally the exact, perfect, identity of initial and final
velocities  will never occur in
reality. But we can still imagine the idealized case and  ask 
ourselves what the nice simple formula looks like. It looks like
this
\cite{me}:

\begin{equation}\label{one}
n ={\alpha\over \pi} \int \int dx_{\mu}{1\over S^2_{i\epsilon}}
dx'_{\mu}={2\alpha\over \pi}\int \int dx_{\mu} {\delta_{\mu \nu}-
{\Delta_\mu
\Delta_\nu\over S^2 }
\over S^2} dx'_{\nu}= {2\alpha\over \pi}\int \int {dx_{\mu}^T
dx_{\mu}^{'T} \over  S^2} 
\end{equation}
In the first writing $S_{i\epsilon}$
is the four-distance between the points $x,x'$ in the following
way:
$S^2_{i\epsilon}= (t-t'+i\epsilon)^2 -({\bf x-x'})^2$. This formula
follows from taking the text book formula for the energy radiated
at frequency $\omega$ and dividing by $\omega$ to get \n
(Planck). This is the only place where   quantum mechanics enters.
The
${i\epsilon}$ arises in making sure the resulting integral is
defined, but we get rid of it in the second writing by some 
manipulations which  lead to the factor 2 and the ``transverse
tangent'' expression in the last writing. The ``transverse
tangent'' means to take the tangent vector $dx$ and to remove the
component of it along the vector $\Delta_{\mu}=(x_{\mu}-x'_{\mu})$
connecting the two points $ x,x'$. This construction cleverly
avoids the threatened singularity as $ x\to x'$, since as two
points
approach each other along a curve the tangents point at each other
and $d{ x}^T\to 0$.
The result  refers of course to the average number of photons
since the number fluctuates---\n is not an integer.

 Such a formula for $n$ is
 quite amusing since it means that any curve satisfying the
``identification'' requirement has a real number belonging to
it--its own
``name''. This is a simple  number, an
invariant, intrinsic 
property of  the curve. The ``identification'' requirement isn't
really
very restrictive since the curve can do almost anything it wants
before it finally goes
back out to infinity parallel (in the four-dimensional sense) to
the direction 
 it came in. ``Almost'' because to avoid that other lurking
danger, the
ultra-violet divergence,  the curve must be smooth and not have any
kinks or jumps. One may verify that Eq~\ref{one} leads to the
correct result for dipole radiation and yields the usual
association between acceleration and radiation.

\section{Euclidean Space}
 
In Minkowski space then, we have a nice
way of associating a  number to a curve. We may now even forget
photons and physics for a moment and wonder if this applies to
ordinary curves, plain 
garden-variety curves in Euclidean space. Indeed, if we take the
last form in Eq~\ref{one} as our starting point, there appears to
be nothing against this. The argument that the transverse tangent
construction is finite as $x\to x'$ still holds. So we write

\begin{equation} \label{two} 
n=- 2\int \int{{\bf dx}^T{\bf
dx'}^T\over ({\bf x-x'})^2}
\end{equation}
 and \n is still an invariant
 dimensionless quantity, an intrinsic property of the \c
--basically
characterizing its ``wiggliness''.
 We  must of course retain our requirement
 concerning the infra-red problem. As the curve goes to $\pm
\infty$ it has the same tangent--becomes the same straight line--
(remember that in four-space  the tangent was the velocity ). But
otherwise, as long as it is smooth, the \c can do anything on its
way from $-\infty$ to $+\infty$. We have dropped the $\alpha/\pi$
since we are now only interested in the purely mathematical
structure, but have kept the (-) sign since this makes things come
out positive.

Fig 1 shows an example, with \n evaluated according to
Eq~\ref{two}. According to  a little Fortran program, the
number
 for this curve is 38.8. Playing with the end points
indicates that the number has essentially gone asymptotic in the
picture.  
For the straight line Eq~\ref{two} gives of course zero.

Furthermore, in  Euclidean space a new opportunity presents itself:
the possibility
of closed curves. In Minkowski space a curve could not go
``backwards'' since it could not have a tangent 
$>45^o$, outside the light cone. But
now with no light cone to cross, there is no reason  not to
consider closed curves. Furthermore, note that for a closed curve
the ``beginning'' and ``end'' of a \tj  are the same, so the
identity of initial and final tangents is automatic.

If all this is true then what is the number of that most basic of
closed \cn s, the  circle? It is: 
\begin{equation}\label{circle}
n_{circle}=2\pi^2\;,
\end{equation} 
as just follows from integrating Eq~\ref{two} directly.
The value is of course a result of the normalization we
chose for our integral, but once we have chosen it, it is the same
for all circles, big ones and small ones; \n depends only the shape
of the curve.
In addition, we also might have the strong suspicion that  among
all plane curves the circle has the smallest \nn. I believe this is
indeed true and hope to present a proof  shortly. 
 To exemplify this, here's a little table for the ellipse with
different eccentricities $\epsilon$.
The first entry, for the circle, is close to $\pi^2=9.87$,   and as
would be expected, the
more eccentric the ellipse the larger \nn.

\begin{table} 
\begin{center}
\begin{tabular}{|l|l|}
\hline
Eccentricity&~~~n/2~~~ \\
\hline
\hline
0&9.83 \\ 
\hline
0.5&9.93\\
\hline
0.7&10.4\\ 
\hline
0.9&13.4\\
\hline
0.95&17.2\\
\hline
0.99&35.2\\
\hline
\end{tabular}
\end{center}
\caption{Numerical evaluation of Eq~\ref{two} for ellipses of
increasing eccentricity.}
\end{table}
 
\section{Inversion and a New Integral Invariant}
Given a curve, what other curves have the same \nn? We might
suspect, because
of the dimensionless character of Eq~\ref{two}, that in
addition to the usual invariances under translation, rotation and
so forth,
that \n is conformally invariant~\cite{inv}. The essential part of 
conformal
invariance is the inversion $x_i\to {x_i\over x^2}$. The inversion
carries closed curves into closed curves except
when the center of inversion is on the curve itself, in which case
the
circle, for example, becomes the straight line going to infinity.
It then
turns out that Eq~\ref{two} is invariant under inversions, except
for these
exceptional cases. In the exceptional cases  one must
universally add $2\pi^2$. This is just right to get Eq~\ref{circle}
when  inverting the straight line, for which \n is zero, to a
circle.

 The fact that
this extra addition is universal-- in this case is the same
 for any closed curve--is suggestive
\cite{gd} of the ``anomaly''. That is we have symmetry breaking,
--here for the inversion-- but in a universal
manner. 
The analysis of this situation leads to the study of a quantity
called $I$~\cite{inv}.
$I$  has  the same value for all curves of a given class, of which
there are four, and represents a novel kind of integral
invariant.  

\begin{figure}
\epsfig{file=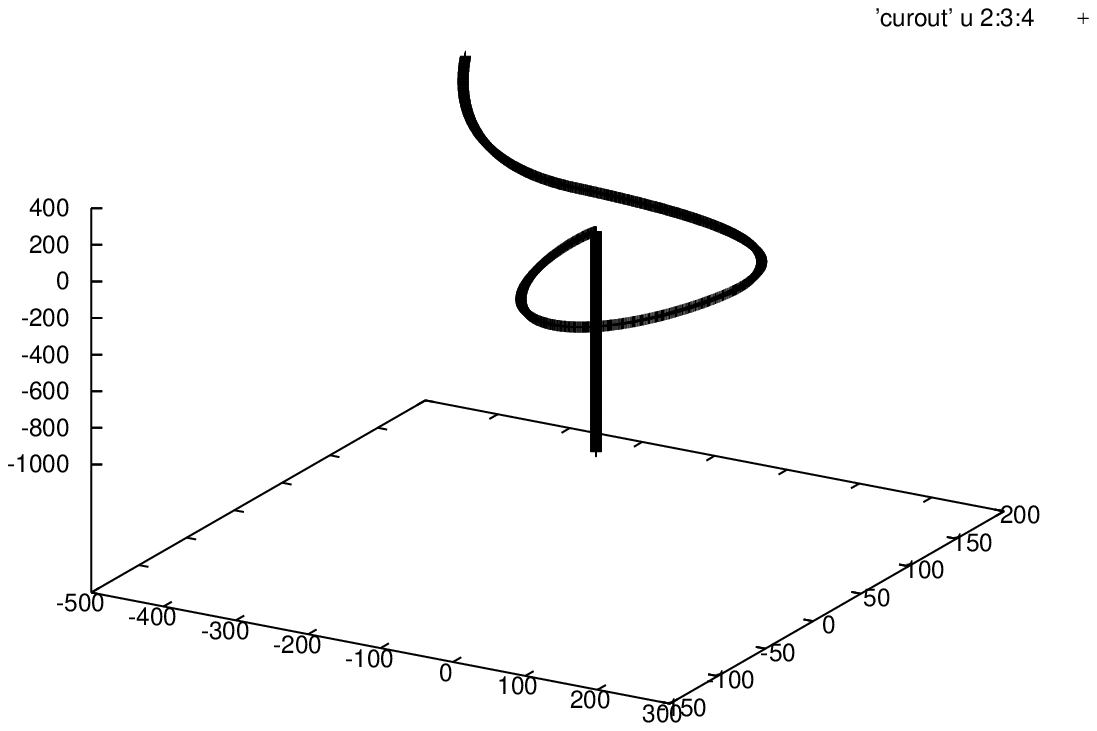,width=0.8\hsize}
\caption{ A \c whose number is 38.8}
\end{figure}

\end{document}